\documentclass[12pt]{article}

\usepackage{epsfig}
\usepackage{amsmath}
\usepackage{amsthm}
\usepackage{amssymb}

\newtheorem{cor}{Corollary}
\newtheorem{pro}{Proposition}

\usepackage{amsopn}

\usepackage{graphicx}

\title{\large\bf Compacton solutions and (non)integrability of nonlinear evolutionary PDEs associated with a chain of prestressed granules}

\author{A. Sergyeyev$^a$, S. Skurativskyi$^b$, V. Vladimirov$^c$\\
$^a$Mathematical Institute, Silesian University in Opava,\\ Na
Rybn\'\i{}\v{c}ku 1, 746 01 Opava,~Czech Republic\\
$^b$Subbotin Institute of Geophysics of NAS of Ukraine,\\ Acad. Palladina Ave. 32, 03142 Kyiv, Ukraine,\\
$^c$Faculty of Applied Mathematics,\\ AGH University of Science
and Technology,\\ Al. Mickiewicza 30, 30059 Krak\'{o}w, Poland\\
E-mails: {\tt Artur.Sergyeyev@math.slu.cz, skurserg@gmail.com},\\ {\tt vsevolod.vladimirov@gmail.com}}

\begin{document}
%\maketitle

\maketitle
\begin{abstract}
%\protect\vspace*{-5mm}
We present the results of study of a nonlinear evolutionary PDE (more precisely, a one-parameter family of PDEs) associated with the chain of pre-stressed granules. The PDE in question supports solitary waves of compression and rarefaction (bright and dark compactons) and can be written in Hamiltonian form. We investigate {\em inter alia} integrability properties of this PDE and its generalized symmetries and conservation laws.\looseness=-1

For the compacton solutions we perform a stability test followed by the numerical study. In particular, we simulate the temporal evolution of a single compacton, and the interactions of compacton pairs. The results of numerical simulations performed for our model are compared with the numerical evolution of corresponding Cauchy data for the discrete model of chain of pre-stressed elastic granules.\looseness=-1



\end{abstract}

%\maketitle

\section{Introduction}
This paper deals with nonlinear evolutionary PDEs associated with dynamics of a one-dimensional chain of pre-stressed granules which arises in quite a number of applications. Since  Ne\-ste\-ren\-ko's pioneering works \cite{Nester_83, Nester_94}  propagation of pulses in such  media has been a subject of a great number of experimental studies and numerical simulations, see \cite{NestLaz_85,Coste_97,Nester_02,Herbold_06,Herbold_07,Ahnert_09, ij, j, Yang_11,lodz2016, vs2018} and references therein.  We consider a nonlinear evolutionary PDE which is derived from the infinite system of ODEs describing the dynamics of one-dimensional chain of elastic bodies  interacting with each other  by means of a nonlinear force. The PDE in question is obtained through the passage to continuum limit  followed by the formal multi-scale decomposition.

The PDE under study turns out to admit a Hamiltonian representation and possess localized traveling wave solutions manifesting some features of solitons. For this reason, it is of interest to investigate its complete integrability. We do this below along with the study of generalized symmetries and conservation laws. We show below that the compacton traveling wave (for traveling waves in general see e.g.\ \cite{Dodd, Valls} and references therein) solutions satisfy the necessary condition for the extremum of a functional associated with the Hamiltonian. Using this we also perform a stability test followed by the numerical study of the compacton solutions. Somewhat surprisingly, numerical simulations show that even in a nonintegrable case the compacton solutions recover their shapes after the collisions, yet the dynamics of interaction slightly differs from that of KdV solitons. In this connection note that compactons, i.e.\ soliton-like solutions with compact support, see \cite{Nester_83,NestLaz_85,Hyman-Rosenau} and references therein, exist for a number of physically relevant models and possess several interesting features making them a subject of intense research, cf.\ e.g.\ \cite{RosenauKmn, Saccomandi, Rosenau_06,  Vodova, CIS, Rosenau17} and references therein.

The paper is organized as follows. In section \ref{vs:sec2} we introduce the continual analog of the granular pre-stressed media with the specific interaction of the adjacent blocks which allows for the description of both the waves of compression and rarefaction.  In section \ref{sec:hs} we present the Hamiltonian structure of the equation in question. In section \ref{sec:cl} we study the conservation laws admitted by the said equation. In section \ref{sec:int} we perform the integrability test that singled out an exceptional integrable case, which is studied in more detail in section \ref{sec:intcase}. In section \ref{vs:sec3} we show that the compacton traveling wave (TW) solutions that satisfy factorized equations also satisfy necessary conditions of extrema for the appropriate Lagrange functionals.  Next we perform  stability tests for compacton solutions based on the approach developed in \cite{Derrick_64, Zakharov_86, Karpman_95}, and show that both dark and bright compactons pass the stability test. The results of qualitative analysis are backed and partly supplemented by the numerical study performed in section ~\ref{vs:sec4}.
We also present the results of numerical simulation of the Cauchy problem for discrete chains and compare the results obtained  with the analogous simulations performed for the continual analogue of these chains. The closing section~\ref{vs:sec5} contains conclusions and discussion.\looseness=-1

\section{Evolutionary PDEs associated with the granular prestressed chains}\label{vs:sec2}

%\noindent
Amazing features of the solitons associated with the celebrated Korteweg--de Vries (KdV) equation, as well as other completely integrable  models \cite{Dodd}, are often ascribed to the existence of higher symmetries and infinite sets of conservation laws, cf.\ e.g.\ \cite{Ibragimov, mikshab, Olver}. However, there exist non-integrable equations possessing localized TW solutions with quite similar behavior. A well-known example of this is provided by the $K(m,n)$ equations \cite{Hyman-Rosenau}:
\begin{equation}\label{Kmn}
K(m, n): u_t+\left( u^m \right)_x+ \left(u^n\right)_{xxx}=0, \qquad m\geq 2,\qquad n\geq 2.
\end{equation}
The members of this hierarchy are not completely integrable at least for generic values of the parameters $m$, $n$, see \cite{RosenauKmn, Vodova} and references therein, and yet possess compactly-supported TW solutions exhibiting solitonic features \cite{Hyman-Rosenau, frutos}.\looseness=-1

The $K(m, n)$ family was introduced in the 1990s as a formal generalization of the KdV hierarchy without referring to its physical context. Earlier V.F. Nesterenko \cite{Nester_83} considered the dynamics of a chain of preloaded granules described by the following ODE system:
\begin{equation}\label{DS_Nest}
\ddot Q_k(t)=F(Q_{k-1}-Q_{k})-F(Q_{k}-Q_{k+1}),   \qquad       k \in \left \{0, \pm 1, \pm 2\dots.   \right\}
\end{equation}
where
$Q_k(t)$ is the displacement of the $k$th granule center-of-mass from its equilibrium position,
\begin{equation}\label{force}
F(z)=A z^n, \qquad n>1.
\end{equation}
%Nesterenko makes an assumption that $B=0$ (see \cite{Nester_2002} for the details), which is validated if the chain is %pre-stressed.
He has described for the first time the formation of localized wave patterns and evolution within this model \cite{Nester_83, Nester_94, NestLaz_85}. In \cite{Nester_83, Nester_94, Nester_02} he presented
the nonlinear evolutionary PDEs being the quasi-continual limits of the discrete models; in this connection cf.\
also \cite{nester_18}.\looseness=-1

The transition to the continual model is achieved via the substitution
\begin{equation}\label{Contanalog}
Q_k(t)=u(t,k\cdot a) \approx u(t,x),
\end{equation}
where $a$ is the average distance between granules.
%We assume in addition that  $m=1,$ $B=\gamma a^{n+1}$, where $\gamma = O(1)$ or zero.
Insert this formula, together with the substitutions
\begin{equation}\label{ContShift}
\begin{array}{rcl}
Q_{k\pm 1}=u(t, x\pm a)&=&\exp(\pm a D_x)u(t, x)\\
&=&\displaystyle\sum_{j=0}^{4} \frac{(\pm a)^j}{j!}\frac{\partial^j}{\partial x^j}u(t, x)
+O\left(a^{5}\right),
\end{array}
\end{equation}
into (\ref{DS_Nest}), and observe that the term of lowest order in $a$ on the right-hand side of (\ref{DS_Nest}) is proportional to $a^{n+1}$. Now expanding the right-hand side of (\ref{DS_Nest}) divided by $a^{n+1}$ into the (formal) Taylor series and then dropping the terms of the order $O(a^{4})$ and higher in this expansion yields from (\ref{DS_Nest}) the equation
%\begin{equation}\label{PDE2_1}
\[
u_{tt}=-C \left\{\left(-u_x  \right)^n+\beta \left(-u_x  \right)^\frac{n-1}{2} \left[\left(-u_x  \right)^\frac{n+1}{2}  \right]_{xx}   \right\}_x,
\]
%\end{equation}
where
\[
C=A a^{n+1}, \qquad \beta=\frac{n a^2}{6(n+1)}.
\]
Differentiating the above equation with respect to $x$ and employing the new variable $S=\left(-u_x  \right)$ corresponding to the strain field, one obtains the Nesterenko equation \cite{Nester_02}:
\begin{equation}\label{EqNest}
S_{tt}=C \left\{S^n+\beta S^\frac{n-1}{2} \left[S^\frac{n+1}{2}  \right]_{xx}   \right\}_{xx}.
\end{equation}
Eq.~(\ref{EqNest}) was derived using only one small parameter corresponding to the long wave approximation.  Thus it can describe the dynamics of ``strongly preloaded media" with dynamic amplitude much smaller than the preload or the dynamics of ``weakly preloaded media" when the dynamic amplitude in the wave is much larger than the preload or even when the preload is equal to zero, in which case the propagation of acoustic waves is impossible (the effect of ``sonic vacuum" \cite{Nester_94}).
As it is shown in \cite{Nester_02}, equation (\ref{EqNest}) possesses a one-parameter family of compacton TW solutions describing the propagation of the waves of compression.

Unfortunately, the compacton solutions supported by (\ref{EqNest}) are unstable.
This can be verified by a direct numerical calculation, substituting in the corresponding difference scheme as Cauchy data known compacton solutions. 

A similar situation occurs in the case of the Boussinesq equation, obtained as a continuum limit of the Fermi--Pasta--Ulam  system of coupled oscillators \cite{Dodd}. As is well known, the  Boussinesq equation possesses unstable soliton-like solutions, and the KdV equation,
supporting the stable uni-directional solitons, is extracted from the Boussinesq equation by means of the asymptotic multi-scale expansion \cite{Dodd}, cf.\ also \cite{burde-s} and references therein.

In this connection it should be also noted that the instability caused by the short wavelengths can be removed using the regularization consisting in replacing the space derivatives of the force by mixed space and time derivatives.  The regularized equation for the case of general power law is nothing but Eq.~(1.110) from \cite{Nester_02}, and its counterpart for a general interaction law is Eq.~(1.156) from \cite{Nester_02}.

It should be further noted that, at least for $n=3/2$, equation (\ref{EqNest}) has stationary compacton solutions which are close to the numerical solutions of the discrete Hertzian chain, see e.g.\ \cite{Ahnert_09}, and the numerical simulations and experiments strongly suggest that the latter solutions are stable. For example, such solutions are generated from various initial conditions on short distances from the disturbed end and propagate in experiments despite disturbances due to inevitable dissipation and violation of periodicity, see e.g.\ experimental results in \cite{Nester_02}.

Another interesting observation is that the conditions for existence of solitary waves in discrete chain \cite{FW} and in the continuum approximation, see Eq.(1.154) at p.~108 in \cite{Nester_02}, are identical and based on the sign of the second derivative of the force, see p.~113 in \cite{Nester_02}.

Our approach to finding a ``proper" compacton-supporting equation is as follows.
%the following one.
We start from the  discrete system (\ref{DS_Nest}) in which the interaction force has the form
\begin{equation}\label{Forceint}
F(z)=A  z^n+B z.
\end{equation}
In addition, we assume that  $B=\gamma a^{n+3}$, where $|\gamma|=O(|A|)$.

The interaction law in (\ref{Forceint}) is a special case of general interaction law that results in long wave equation, Eq.(1.154) at p.~108 in \cite{Nester_02} or its regularized counterpart, Eq.(1.156) in \cite{Nester_02}. The stationary solutions of the said long wave equation are studied in \cite{Nester_02}, where, depending on the behavior of the second derivative of the interaction law, strongly nonlinear compression or rarefaction solitary waves are predicted.

In this connection also note that for small deformations the interaction law in (\ref{Forceint}) is a special case of the situation where the first derivative is nonzero and higher derivatives are zero except for the $n$-th order one which, in conjunction with the discussion in the preceding paragraph implies, cf.\ \cite{Nester_02}, in particular p.110--123, that the linear part of the interaction law is, to an extent, irrelevant for the study of qualitative behavior of stationary solutions.

Inserting  (\ref{Contanalog}), (\ref{ContShift}) into the formula  (\ref{DS_Nest}) and assuming that  the interaction is described by (\ref{Forceint}),  we obtain, up to the terms of the order $O(a^{4})$ and higher in the expansion of the right-hand side of (\ref{DS_Nest}) divided by $a^{n+1}$, cf.\ the discussion after (\ref{ContShift}), the equation
%\begin{equation}\label{PDE2}
\[
u_{tt}=-C \left\{\left(-u_x  \right)^n+\beta \left(-u_x  \right)^\frac{n-1}{2} \left[\left(-u_x  \right)^\frac{n+1}{2}  \right]_{xx}   \right\}_x-\gamma a^{n+3} \left(-u_x  \right)_{x}.
\]
%\end{equation}
Differentiating the above equation with respect to $x$ and introducing the new variable $S=\left(-u_x  \right)$,  we  obtain the following equation:
\begin{equation}\label{EqNest2}
S_{tt}=C \left\{S^n+\beta S^\frac{n-1}{2} \left[S^\frac{n+1}{2}  \right]_{xx}   \right\}_{xx}+\gamma a^{n+3} S_{xx}.
\end{equation}
Now we use a series of scaling transformations. Employing the scaling
$
\tau =\sqrt{\gamma a^{n+3}} t
$
enables us to rewrite the above equation in the form
\[
S_{\tau \tau}=\frac{C}{\gamma a^{n+3}} \left\{S^n+\beta S^\frac{n-1}{2} \left[S^\frac{n+1}{2}  \right]_{xx}   \right\}_{xx}+S_{xx}.
\]
Next, the transformation
$
\bar T=\frac{1}{2} a^q \tau$, $\xi=a^p (x-\tau)$,  $S=a^r W
$
is used. If, for example, we assign  the following values to the parameters
$
q=1$, $ p=-1$, $r=5/n,
$
then the higher-order coefficient $O(a^2)$ will be that of the second derivative with respect to $\bar T$. So, dropping the terms proportional to $O(a^2)$, we obtain, after the integration with respect to $\xi$, the equation:
\[
W_{\bar T}+\frac{A}{\gamma}\left\{W^n+\frac{n}{6 (n+1)} W^\frac{n-1}{2} \left[W^\frac{n+1}{2}  \right]_{\xi \xi}  \right\}_{\xi}=0.
\]
Performing the rescaling and returning to the initial notation
$$
 t=\frac{A}{\gamma} L   \bar T, \qquad x=L \xi,
$$
where $L=\sqrt{\frac{6 (n+1)}{n}}$, we finally obtain the sought-for equation
\begin{equation}\label{PDE3}
W_{T}+\left\{W^n+ W^\frac{n-1}{2} \left[W^\frac{n+1}{2}  \right]_{XX}  \right\}_{X}=0,
\end{equation}
to which we shall hereinafter refer as to the {\em Nesterenko equation}.
Note that Eq.(\ref{PDE3}) appears in \cite{Rosenau_06} (see also \cite{Rosenau17}) as a particular case of the $C_1(m, a+b)$ hierarchy introduced as a generalization of the set of $K(m, n)$ equations.

The description of waves of rarefaction in the case $n=2 k$ requires the following modification of the interaction force:
\begin{equation}\label{Forcerar}
F(z)=-A z^{2 k}+B z
\end{equation}
(for $n=2 k+1$ the formula (\ref{Forceint}) describes automatically both waves of compression and of rarefaction). Applying the above machinery to (\ref{DS_Nest}) with the interaction (\ref{Forcerar}), we obtain, in the same notation, the equation
\begin{equation}\label{PDErar}
W_T-\left\{W^n+ W^\frac{n-1}{2} \left[W^\frac{n+1}{2}  \right]_{XX}  \right\}_{X}=0, \quad n=2 k.
\end{equation}
Thus,  the universal equation describing waves of compression and rarefaction for arbitrary $n \in \mathbb{N}$ can be written  in the form
\begin{equation}\label{PDEComprar}
W_{ T}+\left[\mathrm{sgn}(W)\right]^{n+1} \left\{W^n+ W^\frac{n-1}{2} \left[W^\frac{n+1}{2}  \right]_{XX}  \right\}_{X}=0.
\end{equation}

In closing note that equations  (\ref{PDE3}), (\ref{PDErar}) and (\ref{PDEComprar}) are obtained by formal application of the multiscale decomposition method which cannot be substantiated in our case because of negativity of the index $p$, cf.\ \cite{Rosenau_03} where this problem is discussed in a more general fashion. Further study of these equations is justified  by the fact that  they possess a  set of compacton  solutions possessing interesting dynamical features. As will be shown below, these solutions describe well enough propagation of short impulses in the chain of pre-stressed blocks.

%%%%  A_Serg   %%%%%%%%%%%%%

\section{Hamiltonian structure for the Nesterenko equation}\label{sec:hs}

%With a slight abuse of notation, we return to the small letters and rewrite equation (\ref{PDE3}) in the following form:
%which was derived in ?? using an asymptotic technique similar to that from \cite{bus},

Now return to (\ref{PDE3}) which we now write in the manifestly evolutionary form, that is,
\begin{equation}\label{nest}
W_T=-\left(W^n+W^{(n-1)/2} \left[W^{(n+1)/2}  \right ]_{XX}  \right)_X
\end{equation}
%   where $n$ is a parameter.
%assuming that $\beta>0$ and $\gamma\neq 0$.
Note that for $n=-1$ this equation boils down to a quasilinear first-order equation $W_T=(W^{-1})_X$ which is obviously integrable, and for $n=1$ equation (\ref{nest}) becomes linear.

Equation (\ref{nest}) can be written (cf.\ \cite{Rosenau17}) as
\begin{equation}\label{n}
W_T=D_X\delta \mathcal{H}_{\mathrm{Nest}}/\delta W\equiv F.
\end{equation}
%wh
Thus, (\ref{n}) is written in Hamiltonian form with the Hamiltonian $\mathcal{H}_{\mathrm{Nest}}$ and the Hamiltonian structure $\mathfrak{P}_0=D_X$.

This implies, in particular, that
%$\mathfrak{P}_0$ maps cosymmetries to symmetries, so to any local cosymmetry, and, in particular,
to any nontrivial local conserved density of (\ref{n}) there corresponds a (generalized, but not necessarily genuinely generalized (see the definition below), and possibly trivial) symmetry of (\ref{n}).

Here $\delta/\delta W$ is the variational derivative (see below for details) and $\mathcal{H}_{\mathrm{Nest}}=\displaystyle\int h_{\mathrm{Nest}} dX$ with the density
%, for $n\neq -1$
\begin{equation}\label{h}
h_{\mathrm{Nest}}=\left\{\begin{array}{l}\displaystyle\left(\frac14 (n+1)W^{n-1}W_X^2-W^{n+1}/(n+1)\right)\quad\mbox{for $n\neq-1$},\\[5mm] \displaystyle\ln |W|\quad\mbox{for $n=-1$.}\end{array}\right.
\end{equation}
Here and below the integrals are understood in the sense of formal calculus of variations, see e.g.\ \cite{Olver, Dorfman}.
%and $\mathcal{H}=-\displaystyle\int dx \ln |u|$ for $n=-1$.
%and $n$ is a real number.
Here we put, cf.\ \cite{mikshab, Olver, Dorfman}, $W_j=\partial^j
 W/\partial X^j$, $j=1,2,\dots$, $W_0\equiv W$, and define \cite{Ibragimov, mikshab, Olver, Dorfman} the total derivatives
\begin{equation}\label{td}
D_X=\displaystyle\frac{\partial}{\partial X}+\sum\limits_{j=0}^\infty W_{j+1}\displaystyle\frac{\partial}{\partial W_j},\quad
D_T=\displaystyle\frac{\partial}{\partial T}+\sum\limits_{j=0}^\infty D_X^{j}(F)\displaystyle\frac{\partial}{\partial W_j}.
\end{equation}
The variational derivative of a functional $\mathcal{F}=\displaystyle\int f(X,T,W,W_1,\dots,W_k)dX$ has the form
\begin{equation}\label{vd}
\frac{\delta\mathcal{F}}{\delta W}=\sum \limits_{j=0}^\infty (-D_X)^{j}\left(\displaystyle\frac{\partial f}{\partial W_j}\right).
\end{equation}
For any $f=f(X,T,W,\dots,W_k)$ we also define, cf.\ e.g.\ \cite{mikshab, Olver}, its linearization
\[
f_*=\sum\limits_{j=0}^k \displaystyle\frac{\partial f}{\partial W_j} D_X^j.
\]

\section{Conservation laws}\label{sec:cl}
Recall, cf.\ e.g.\ \cite{Ibragimov, mikshab, Olver, Dorfman, bp, ps, scg, fer} and references therein, that a local conservation law for (\ref{nest}) is, roughly speaking, a relation of the form
\begin{equation}\label{c}
D_T(\rho)=D_X(\sigma),
\end{equation}
where $\rho=\rho(X,T,W,W_1,\dots,W_r)$ and $\sigma=\sigma(X,T,W,W_1,\dots,W_s)$, which holds by virtue of (\ref{nest}).
Here $\rho$ and $\sigma$ are called a (conserved) density and the flux of our conservation law.

Also recall, cf.\ e.g.\ \cite{ps}, that a conservation law (\ref{c}) is called nontrivial if there exists no function %
$\zeta(X,T,W,W_1,\dots,W_{q})$ such that $\rho=D_X\zeta$, i.e., $\rho\not\in\mathrm{Im} D_X$.

It is well known, see e.g.\ \cite{Olver, Dorfman}, that a necessary and sufficient condition for a function $f=f(X,T,W,W_1,\dots,W_r)$ to not belong to the image of $D_X$ is $E_W f\neq 0$, where $E_W$ is the Euler operator
\[
E_W=\sum \limits_{j=0}^\infty (-D_X)^{j}\circ \displaystyle\frac{\partial}{\partial W_j}.
\]

Hence $\rho$ is a conserved density for (\ref{nest}) if and only if $E_W D_T(\rho)=0$, and this density is nontrivial if and only if $E_W \rho\neq 0$.

It is readily checked that we have the following
%, in particular,
%that
\begin{pro}\label{cl1}
For any $n$ equation
(\ref{nest}) admits the following three conserved densities:
\begin{equation}\label{d}
\rho_0=W,\quad \rho_1=W^2/2,\quad \rho_2=h_{\mathrm{Nest}}.
\end{equation}
%and for $n=-1$ the density $\rho_2$ should read $-\ln|u|$.
For $n=0$ we have an extra density
%in the above class, i.e., of the form $\rho=\rho(X,T,W,W_X,\dots,W_{XXXXX})$,
\begin{equation}\label{d0}
\rho_3=X^2 W W_X -T W_X^2/W.
\end{equation}
Moreover, for $n\neq 0,1,-1,-2$ (resp.\ for $n=0$) the densities (\ref{d}) (resp.\ (\ref{d} and (\ref{d0})) exhaust, modulo the addition of trivial ones,  the linearly independent conserved densities of order up to five, i.e., of the form $\rho=\rho(X,T,W,W_X,\dots,W_{XXXXX})$.
\end{pro}
It is very likely that for $n\neq 1,-1,-2$ no local conserved densities of order greater than five (of course, again modulo trivial ones) exist at all in view of nonintegrability of (\ref{nest}) for $n\neq 1,-1,-2$ as discussed below.

Recall that $\rho_2$ is the density of the Hamiltonian $\mathcal{H}_{\mathrm{Nest}}$ for (\ref{nest}) with respect to the Hamiltonian structure $\mathfrak{P}_0=D_X$. To the functional $\mathcal{C}=\displaystyle\int W dX$ there corresponds a trivial symmetry, i.e., a symmetry with zero characteristic, as $D_X \delta\mathcal{C}/\delta W=0$, so $\mathcal{C}$ is a Casimir functional for $\mathfrak{P}_0$. To the functional $\mathcal{P}=\frac12\displaystyle\int W^2dX$ there corresponds a symmetry with the characteristic $W_X=D_X\delta\mathcal{P}/\delta W$, that is, $X$-translation, and to $\mathcal{H}_{\mathrm{Nest}}$ there corresponds a symmetry with the characteristic equal to the r.h.s.\ $F$ of (\ref{n}), i.e., the time translation symmetry.

%If furthermore $n\neq 0,1,-1,-2$, then direct verification shows that (\ref{d}) are the only conserved densities (modulo trivial ones) for (\ref{nest}) %in the class of local conserved densities of order up to five, i.e., of the form $\rho=\rho(X,T,W,W_X,\dots,\allowbreak W_{XXXXX})$.

For $n=0$
%we have an extra density in the above class, i.e., of the form $\rho=\rho(X,T,W,W_X,\dots,W_{XXXXX})$,
%\begin{equation}
%\rho_3=X^2 W W_x -t W_X^2/W,
%\end{equation}
%and
to the conserved functional $\mathcal{H}_3=\displaystyle\int\rho_3 dX$ there corresponds a scaling symmetry with the characteristic $4T F+2 X W_X+2W=D_X\delta\mathcal{H}_3/\delta W$. Again, it is very likely that $\rho_i$, $i=0,\dots,3$, are the only local conserved densities (modulo trivial ones) for (\ref{nest}) with $n=0$ in view of nonintegrability of this special case of (\ref{nest}).\looseness=-1

\section{Integrability}\label{sec:int}
Integrable equations of the form (\ref{n}) with the Hamiltonian of general form $\mathcal{H}=\displaystyle\int dX h(W,W_X)$
%with a density of general form
where the density $h=h(W,W_X)$ is such that $\partial^2 h/\partial  W_X^2\neq 0$ were classified (modulo point transformations leaving $T$ invariant) in \cite{ms}. Note that in \cite{mikshab, ms} and references therein integrability of an evolution equation
\begin{equation}\label{ee}
W_T=K(X,W,W_X,\dots,\partial^k W/\partial X^k)
\end{equation}
with $k\geq 2$ means existence of an infinite hierarchy of generalized symmetries of increasing orders which do not depend explicitly on $T$. In order to avoid ambiguity we shall, following the common usage, refer below to this kind of integrability as to the {\em symmetry integrability}.\looseness=-1

Recall, cf.\ e.g.\ \cite{Ibragimov, mikshab, Olver, Dorfman}, that a {\em generalized symmetry} of order $r$ for (\ref{ee}) is\footnote{For the sake of simplicity and without loss of generality we identify here a generalized symmetry with its characteristic.}
a function $G=G(X,T,W,W_1,\dots,\allowbreak W_r)$ such that $\partial G/\partial W_r\neq 0$ and
\begin{equation}\label{symm}
D_T(G)=K_*(G),
\end{equation}
where now $D_T=\displaystyle\frac{\partial}{\partial T}+\sum\limits_{j=0}^\infty D_X^{j}(K)\displaystyle\frac{\partial}{\partial W_j}$.

Such a symmetry $G$ is known as {\em genuinely generalized} if it cannot be written in the form $G=c(T)K+b(X,T,W,W_X)$ for some functions $b$ and $c$, that is, it is not equivalent to a point or contact symmetry. As far as point symmetries of the equations studied in the present paper, and, more broadly, of $\mathcal{C}_1(m,a,b)$ equations (see e.g.\ \cite{Rosenau_06, Rosenau17}), cf.\ e.g.\ \cite{Tracina} and references therein.\looseness=-1

Thus, symmetry integrability of (\ref{ee})
%in the above sense
means existence of an infinite hierarchy of generalized symmetries of the form $G_i(X,W,W_1,\dots,W_{r_i})$ of increasing orders $r_i$.

Now turn to comparison of the density $h_{\mathrm{Nest}}$ of our Hamiltonian and the densities $h$ found in \cite{ms} for which the equation $W_T=D_X(\delta\mathcal{H}/\delta W)$
with the general Hamiltonian $\mathcal{H}=\displaystyle\int h(W,W_X)dX$ is symmetry integrable. % in the above sense.

\begin{pro}\label{sym-int}
The only symmetry integrable case of (\ref{nest}) which is genuinely nonlinear and genuinely of third order is that of $n=-2$.
\end{pro}

\noindent{\em Proof.}
It is not difficult to observe (cf.\ e.g.\ \cite{vps}) that using point transformations leaving $t$ invariant the density $h_{\mathrm{Nest}}$ of our Hamiltonian for $n\neq -1$ can, if at all, only be transformed into just one case from \cite{ms}, namely, equation (2.1) in \cite{ms}, that is,
\begin{equation}\label{hms}
h=W_X^2/(2 f^3)-P/f,
\end{equation}
where $f=c_0+c_1 W +c_2 W^2$, $P=\sum\limits_{i=0}^4 d_i W^i$, and $c_i$ and $d_i$ are arbitrary constants.

Moreover, it is clear that in our case $f$ should actually be a monomial: $f=cW^\alpha$, $\alpha=0,1,2$.

Upon comparing the coefficients at $W_X^2$ in (\ref{hms}) and (\ref{h}) modulo an obvious rescaling of $W$, we see that all values of $n$ for which (\ref{nest}) could be integrable should satisfy $n-1=0,-3,-6$. The case of $n=1$ is trivially integrable, as then (\ref{nest}) is just a linear equation,
so we are left with two possibilities $n=-2$ and $n=-5$ corresponding to $\alpha=1$ and $\alpha=2$.

Now upon inspecting the remaining terms in $h_{\mathrm{Nest}}$ and in (\ref{hms}) we readily conclude that the polynomial $P$ should also reduce to a single monomial: $P=d W^\beta$, where $\beta=0,1,2,3,4$, so we have a system $n-1=-3\alpha$ and $n+1=\beta-\alpha$, where $\alpha=1,2$ and $\beta=0,1,2,3,4$. An obvious corollary of this system is $-3\alpha+2=\beta-\alpha$, whence $\beta=2(1-\alpha)$.
However, $\beta\geq 0$ by assumption, so the case of $n=-5$, when $\alpha=2$ and we should have $\beta=-2$, is not integrable.

Thus, the only integrable case of (\ref{nest}) which is genuinely nonlinear and genuinely of third order is that of $n=-2$, and the result follows. $\Box$.

Recall that for $n=-1$ equation (\ref{nest}) degenerates and becomes a first order quasilinear equation whose general solution can be found, see above, and for $n=1$ equation (\ref{nest}) is just linear.

%Denote by $F$ the right-hand side of
%(\ref{nest}).
In fact, the %above nonintegrability
result of Proposition~\ref{sym-int} can be further strengthened so that absence of any generalized symmetries, rather than just those that do not depend explicitly on $T$, can be established.

To this end consider, following \cite{mikshab}, the so-called canonical density $\rho_{-1}=(\partial F/\partial W_{XXX})^{-1/3}$.
% and $\rho_0=(\partial F/\partial W_{XX})/(\partial F/\partial W_{XXX})$,
%
It is readily checked that $E_W D_T(\rho_{-1})\neq 0$ for $n\neq -1,-2,-5,1$.
Hence for $n\neq -1,-2,-5,1$ we have $D_T(\rho_{-1})\not\in\mathrm{Im}\ D_X$, and thus $\rho_{-1}$ is not a density of
a local conservation law for (\ref{nest}).

In turn, by virtue of the results from \cite{as} this immediately implies %that
\begin{pro}\label{sym-pro}
Equation (\ref{nest})
for $n\neq 1,-1,-2,-5$ has no generalized symmetries of order greater than three.
\end{pro}
In other words, Proposition~\ref{sym-pro} means that
for $n\neq 1,-1,-2,-5$ any solution $G=G(X,T,W,W_1,\dots,W_r)$
of %(\ref{symm})
the equation
\begin{equation}\label{sym}
D_T (G)=F_*(G),
\end{equation}
where $D_T$ and $F$ are given in (\ref{td}) and (\ref{n}),
%i.e., a generalized symmetry for (\ref{nest}) with $n\neq -1,-2,-5,1$
in fact depends at most on $X,T,W,W_X,W_{XX},W_{XXX}$.
%Moreover, it can be shown that (\ref{nest}) for $n\neq 1,-1,-2,-5$ admits no genuinely generalized symmetries, including those with explicit dependence %on $T$.

%Also, it is easily seen that $\delta D_t(\rho_{0})/\delta u\neq 0$ for $n\neq -1,1$, so $\rho_{-1}$ is not a density of
%a local conservation law for (\ref{nest}) for $n\neq -1,1$.

This implies that (\ref{nest}) for $n\neq -1,-2,-5,1$  admits no genuinely generalized symmetries,
and hence (\ref{nest}) for $n\neq -1,-2,-5,1$ is unlikely to be integrable in any reasonable sense, cf.\  \cite{mikshab}.

Leaving aside the degenerate cases of $n=\pm 1$, turn to the remaining two special cases: $n=-2$ and $n=-5$. We believe that using the technique similar to that of \cite{sv} (cf.\ also \cite{Vodova, v}) it can be shown that in the case of $n=-5$ equation (\ref{nest}) admits no genuinely generalized symmetries, including those with explicit dependence on $T$ and not just the time-independent ones whose nonexistence follows from the above comparison of (\ref{h}) with (\ref{hms}), so we are left with just one integrable case of $n=-2$ which we discuss below.\looseness=-1

\section{Nesterenko equation for $n=-2$: integrability and beyond}\label{sec:intcase}
%Lax pair and bihamiltonian structure}
The following result is readily checked by straightforward computation:
\begin{pro}\label{intcase}
For $n=-2$ equation (\ref{nest})
%is integrable, as it
has a Lax pair of the form
\begin{equation}\label{lax}
\psi_{XX}=(1+W^2\lambda)\psi,\quad
\psi_T=\frac{2}{W^3}\psi_{XXX}-\frac{3W_X}{W^4}\psi_{XX}+\frac{2}{W^3}\psi_X-\frac{3W_X}{W^4}\psi
\end{equation}
and %also
admits a recursion operator %of the form
\begin{equation}\label{ro}
\begin{array}{rcl}
\mathfrak{R}&=&\displaystyle\frac{1}{W^2} D_X^2-\frac{3 W_X}{W^3}D_X+\frac{(4 W^2+6 W_X^2-3 W_{XX})}{W^4}\\[5mm]
&&-2\left(W^{-2}+W^{-3/2} \left[W^{-1/2}  \right ]_{XX}  \right)_X D_X^{-1}
\end{array}
\end{equation}
\end{pro}
The recursion operator (\ref{ro}) can be found e.g.\ using the technique from \cite{mas} (cf.\ also \cite{sc}).
%where $u_t$ stands for the right-hand side of (\ref{nest}) with $n=-2$.
%
Also note that upon passing to a new dependent variable equal to a square of $W$ the first equation of (\ref{lax}) can be identified with a special case of the eigenvalue problem related to the extended Harry Dym systems, see e.g.\ \cite{MP} and references therein.\looseness=-1

Equation (\ref{nest}) for $n=-2$ also admits a second local Hamiltonian operator $\mathfrak{P}_1=\mathfrak{R}\circ D_x$, that is,
\[
\begin{array}{rcl}
\mathfrak{P}_1&=&\displaystyle\frac{1}{W^2} D_X^3-\frac{3 W_X}{W^3}D_X^2+\frac{(4 W^2+6 W_X^2-3 W_{XX})}{W^4}D_X\\[5mm]
&&-2\left(W^{-2}+W^{-3/2} \left[W^{-1/2}  \right ]_{XX}  \right)_X
\end{array}
%%1/u^2 D_x^3-3 u_x/u^3 D_x^2+(-3 u u_{xx}+6 u_x^2+4 u^2)/u^4 D_x-1/u^5 (4 u^2 u_x +6 u_x^3-6 u u_x u_{xx}+u^2 u_{xxx}),
\]
which is compatible with $\mathfrak{P}_0=D_X$, so the recursion operator $\mathfrak{R}$ is hereditary and equation (\ref{nest}) for $n=-2$ can be written, in addition to (\ref{n}), in the {\em second} Hamiltonian form as
\begin{equation}\label{ham}
W_T=\mathfrak{P}_0(\delta\tilde h/\delta W),
\end{equation}
where $\tilde{h}=W/2$.

Thus, we have the following
\begin{pro}\label{biham}
Equation (\ref{nest}) for $n=-2$ is an integrable bihamiltonian system with two local Hamiltonian operators $\mathfrak{P}_0$ and $\mathfrak{P}_1$ and two local Hamiltonian representations (\ref{n}) and (\ref{ham}).
\end{pro}

Using general theory of bihamiltonian systems (see e.g.\ \cite[Ch. 7]{Olver} and \cite{olver-bih}), we also readily obtain
\begin{cor}\label{biham-c}
Equation (\ref{nest}) for $n=-2$ possesses an infinite hierarchy of commuting generalized symmetries of the form $\mathfrak{R}^k W_X$, $k=0,1,2,\dots$ and an infinite hierarchy of local conservation laws whose densities $h_j$ are generated recursively through the relations
\[
\mathfrak{P}_0(\delta h_{j+1}/\delta W)=\mathfrak{P}_1(\delta h_j/\delta W),
\]
where $j=0,1,2,\dots$ and $h_0=W/2$, and of associated integrals of motion $\mathcal{H}_j=\int h_j dX$ in involution with respect to the two Poisson brackets associated with $\mathfrak{P}_0$ and $\mathfrak{P}_1$.
\end{cor}

The fact that the generalized symmetries $\mathfrak{R}^k W_X$ and the conserved densities $h_k$ for $k=0,1,2,\dots$ do not involve any nonlocal terms can be established using the results of \cite{sw} or \cite{sro} (cf.\ also \cite{sc}).

As we have already pointed out above, up to a suitable rescaling of $T$ and obvious change of notation equation (\ref{n}) for $n=-2$ is a special case of equation (2.1c) in \cite{ms}, and hence can be transformed into a special case of the well-known $S$-integrable Calogero--Degasperis--Fokas \cite{cd, f} equation in the manner described therein. % \cite{ms}.

Namely, pass first to the potential form of (\ref{n}) with $n=-2$,
\[
%\begin{equation}\label{eqv}
V_T=-\frac{V_{XXX}}{2 V_X^3}+\frac{3 V_{XX}^2}{4 V_X^4}+\frac{1}{V_X^2},
%\end{equation}
\]
related to (\ref{nest}) through the differential substitution $W=V_X$.

The subsequent hodograph transformation interchanging $X$ and $V$
turns the above equation into a constant separant equation
\[
V_T=-\frac{V_{XXX}}{2}+\frac{3 V_X V_{XX}}{2 V}-\frac{3 V_X (4+V_X^2)}{4 V^2},
\]
or, upon a suitable rescaling of $T$,
\begin{equation}\label{eqv}
V_T=V_{XXX}-\frac{3 V_X V_{XX}}{V}+\frac{3 V_X (4+V_X^2)}{V^2}.
\end{equation}

Finally, putting $V=\exp(U/2)$ turns (\ref{eqv}) into a special case of the Calogero--Degasperis--Fokas \cite{cd, f} equation, {\em viz.},
\begin{equation}\label{cdf}
U_T=U_{XXX}-\frac18 U_{X}^3+6 U_X\exp(-U).
\end{equation}

%%%%%%%%%%%   END A_Serg   %%%%%%%%%%%%%
%%%%%%%%%%%   END A_Serg   %%%%%%%%%%%%%
%%%%%%%%%%%   END A_Serg   %%%%%%%%%%%%%
%%%%%%%%%%%   END A_Serg   %%%%%%%%%%%%%
%%%%%%%%%%%   END A_Serg   %%%%%%%%%%%%%

\section{Compacton solutions and stability tests}\label{vs:sec3}

%\noindent
%In order to stress that the variables are scaled with respect to the original ones, we  return below to the notation from the end of Section %\ref{vs:sec2}.
Consider the pair of equations (\ref{PDE3}), (\ref{PDErar}), which can be represented by the single expression
\begin{equation}\label{Comprar}
W_{ T}+\epsilon \left\{W^n+ W^\frac{n-1}{2} \left[W^\frac{n+1}{2}  \right]_{XX}  \right\}_{X}=0, \quad \epsilon=\pm 1.
\end{equation}
As we are interested in the traveling wave (TW) solutions $W=W(z)\equiv W(X-c T)$,  it is convenient
%instructive
to
%make a passage
pass to the TW coordinates  $T\to T$, $X\to z=X-c T$. This change of variables yields from (\ref{Comprar}) the equation
\begin{equation}\label{PDE3C}
W_{T}-c W_{ z}+\epsilon \left\{W^n+ W^\frac{n-1}{2} \left[W^\frac{n+1}{2}  \right]_{zz}  \right\}_{z}=0.
\end{equation}

It is easy to check that
equation (\ref{PDE3C})
admits a Hamiltonian formulation
\begin{equation}\label{Hamrepr2}
%\frac{\partial}{\partial T}
W_T = D_z
{\delta \left( \epsilon \mathcal{H}_{\mathrm{Nest}} +c \mathcal{P}\right)}/{\delta W},
\end{equation}
where now
%Recall that
\[
%H=\int\left[\frac{n+1}{4}W^{n-1}  W_z^{2}-\frac{1}{n+1} W^{n+1}\right] d z,
%\qquad
 \mathcal{H}_{\mathrm{Nest}}=\int  h_{\mathrm{Nest}} dz,\qquad \mathcal{P}=\int \frac{1}{2} W^2 dz,
\]
and
\[
h_{\mathrm{Nest}}=\left\{\begin{array}{l}\displaystyle\left(\frac14 (n+1)W^{n-1}W_z^2-W^{n+1}/(n+1)\right)\quad\mbox{for $n\neq-1$},\\[5mm] \displaystyle\ln |W|\quad\mbox{for $n=-1$.}\end{array}\right.
\]

The above formulation up to the coefficient $\epsilon$ follows directly
from the Hamiltonian form (cf.\ (\ref{n}))  of equation (\ref{PDE3}) after the change of coordinates.
Recall that both functionals $\mathcal{H}_{\mathrm{Nest}}$ and $\mathcal{P}$ are conserved in time.

Now consider the following functions:
\begin{equation}\label{comp2a}
W_c^\epsilon(z)=\epsilon W_c(z)=\begin{cases}\epsilon  M \cos^{\gamma} \left(K z\right), & \mathrm{if}\quad |K z|<\frac{\pi}{2}, \\
0 & \quad \mathrm{otherwise},\\
\end{cases}
\end{equation}
where $\epsilon = \pm 1$,
\[
 M=\left[\frac{c (n+1)}{2}  \right]^{\frac{1}{n-1}}, \qquad K=\frac{n-1}{n+1},  \qquad \gamma=\frac{2}{n-1}.
\]
It is readily checked that we have the following
\begin{pro}\label{gs}
If $n=2 k+1$, $k \in \mathbb{N}$, then the functions $W_c^\pm(z)$ are weak solutions to the equation
\begin{equation}\label{variat2D}
\delta \left(\mathcal{H}_{\mathrm{Nest}} +c \mathcal{P}\right)/\delta W|_{W=W_c^\pm} =0.
\end{equation}
If $n=2 k$, $k \in \mathbb{N}$, then the functions $W_c^{\pm 1}(z)$ are weak solutions to the equation
\begin{equation}\label{variateps}
\delta \left( \pm \mathcal{H}_{\mathrm{Nest}} +c \mathcal{P}\right)/\delta W|_{W=W_c^{\pm 1}} =0.
\end{equation}
\end{pro}

So, the TW solutions  (\ref{comp2a}) are  the critical points of either the Lagrange functional $\Lambda=\mathcal{H}_{\mathrm{Nest}} +\beta \mathcal{P}$ (the case of $n=2 k+1$) or
$\Lambda^\epsilon=\epsilon \mathcal{H}_{\mathrm{Nest}} +\beta\mathcal{P}$ (the case of $n=2 k$) with the common Lagrange multiplier $\beta=c$.
As is well known, necessary and sufficient
condition for $\Lambda$ (resp.\ $\Lambda^\epsilon$) to attain the minimum on the compacton solutions can be stated in terms of the positivity of the
second variation of the corresponding functional, which, in turn, guarantees the orbital stability of the TW solution \cite{KaPromis}. Here we do not touch upon the problem of strict estimating of the signs of the second variations. We follow instead the approach suggested in \cite{Derrick_64,Zakharov_86,Karpman_95}, which enables us to test the {\em possibility} of appearance of the local minimum  on selected sets of perturbations of TW solutions.\looseness=-1

Consider the following family of perturbations
\begin{equation}\label{scaling1A}
W_c^\epsilon(z) \rightarrow \lambda^\alpha W_c^\epsilon(\lambda z).
\end{equation}
Upon choosing $\alpha=1/2$ we obtain
\begin{equation}\label{Qinv}
\mathcal{P}[\lambda]=\frac{1}{2}\int_{-\pi/2}^{\pi/2}{\left[\lambda^\frac{1}{2} W_c^\epsilon(\lambda z)\right]^2 d z}=\mathcal{P}[1].
\end{equation}
Thus, for this choice $\mathcal{P}[\lambda]$ keeps its unperturbed value.
By imposing this condition we reject ``fake'' perturbations associated with the translational symmetry $T_\delta\left[W_c^\epsilon(z)\right]=W_c^\epsilon(z+\delta)$. Indeed, since equations (\ref{variat2D}), (\ref{variateps}) are invariant under the shift $z \to  z+\delta$,  $T_\delta W_c^\epsilon(z)$ belongs to the set of solutions as well, while formally the transformation $W_c^\epsilon(z) \rightarrow W_c^\epsilon(z+\delta)$ can be treated as a perturbation. In order to exclude the perturbations of this sort, the orthogonality condition  is imposed. Introducing the representation for the perturbed solution
\[
W_c^\epsilon(z)[\lambda]=W_c^\epsilon(z)+v(z, \lambda),
\]
and using the condition (\ref{Qinv}), we find
\[
0=\mathcal{P}[\lambda]-\mathcal{P}[1]=\int_{-\pi/(2 K)}^{\pi/(2 K)} W_c^\epsilon(z) v(z, \lambda) d z+O\left(||v(z, \lambda)||^2\right),
\]
so if $\mathcal{P}$ is independent of $\lambda$, then, up to $O\left(||v(z, \lambda)||^2\right)$ the perturbation created by the scaling transformation is orthogonal to the TW solution.

For $\alpha=1/2$ and $n\in\mathbb{N}$, we arrive at the following functions to be tested:
\begin{equation}\label{flam1Q}
\Lambda^\nu[\lambda]=(\nu \mathcal{H}_{\mathrm{Nest}} +c \mathcal{P})[\lambda]=\nu \left\{\lambda^{\frac{n+3}{2}}  I_{n}^\epsilon-\lambda^{\frac{n-1}{2}}  J_{n}^\epsilon\right\}+c \mathcal{P},
\end{equation}
where
\[
 I_{n}^\epsilon=\frac{n+1}{4}\int_{-\pi/(2 K)}^{\pi/(2 K)} \left[W_c^\epsilon\right]^{n-1} \left[\left(W_c^\epsilon\right)_z\right]^2 d z,\qquad
 J_{n}^\epsilon=\frac{1}{n+1}\int_{-\pi/(2 K)}^{\pi/(2 K)} \left[W_c^\epsilon\right]^{n+1} d z,
\]
\[
\nu=\epsilon^{n+1}=\begin{cases} +1  & \mathrm{if}\quad n=2 k+1, \\
\quad \epsilon  &  \mathrm{if}\quad n=2 k. \\
\end{cases}
\]

If the functional $\Lambda^\nu=\nu \mathcal{H}_{\mathrm{Nest}} +c \mathcal{P}$ attains the extremal value on the compacton solution, then the function $\Lambda^\nu[\lambda]$ has the corresponding extremum in the point $\lambda=1$. The verification of this property is employed as a test.

A necessary condition for the extremum
$
\frac{d}{d \lambda}\Lambda^\nu[\lambda]\Bigl|_{\lambda=1}=0
$
gives us the equality
\begin{equation}\label{In2}
 I_n^\epsilon=\frac{n-1}{n+3} J_n^\epsilon.
\end{equation}
Using (\ref{In2}), we obtain the estimate
\[
\frac{d^2}{d \lambda^2}\Lambda^\nu[\lambda]\Bigl|_{\lambda=1}=\nu (n-1) J_n^\epsilon =
\frac{n-1}{n+1} \epsilon^{2(n+1)} \int \left[W_c^\epsilon\right]^{n+1}(z) d z > 0,
\]
which is valid for both $n=2 k+1$ and $n=2 k$. Thus, the generalized solutions (\ref{comp2a})  pass the test for stability, and we can state the following
%conjecture

%\begin{conj*}
\noindent{\bf Conjecture.}
{\em For $n>1$ weak solutions (\ref{comp2a}) provide minima of the functional $\Lambda^\nu$.}
%\end{conj}

Further information about the properties of the compacton solutions is provided by the numerical simulations discussed below.

\section{Numerical simulations for dynamics of compactons}\label{vs:sec4}

%\noindent
The dynamics of  solitary waves is studied by means of
direct numerical simulation based on the finite-difference scheme.

\begin{figure}[h]
\begin{center}
%\includegraphics[width=5.5 cm, height=5
%cm]{ris_comp1.eps} \hspace{1.5 cm}
%\includegraphics[width=5.5 cm, height=5 cm]{ris_ruh_1.eps}\\
%a) \hspace{6cm} b) \\
\includegraphics[width=5.5cm, height=5cm]{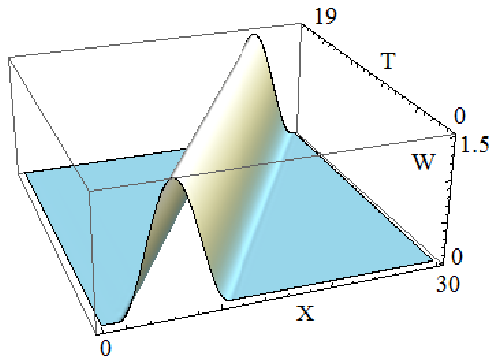}\hspace{0.5cm}
\includegraphics[width=0.5\linewidth]{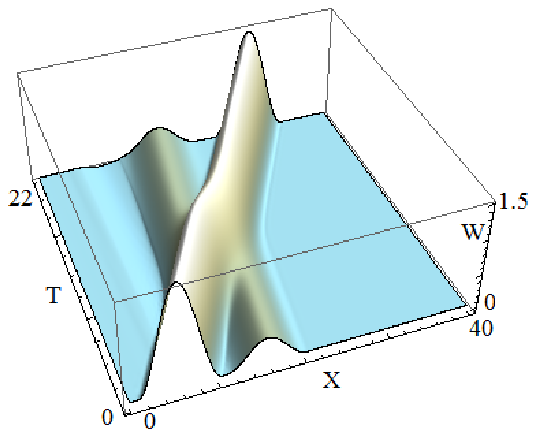}\\
a) \hspace{6cm} b)\\
\end{center}
\caption{
Numerical evolution of a single compacton solution of  Eq. (\ref{NestNeq3}) characterized by the velocity
$c=1$  (a)
 and
%. Numerical evolution of
a pair of  compacton solutions characterized by the velocities $c=1$ and  $c=1/4$ (b), respectively.
}\label{fig:1}
\end{figure}

\begin{figure}
\begin{center}
\includegraphics[width=5.5 cm, height=5
cm]{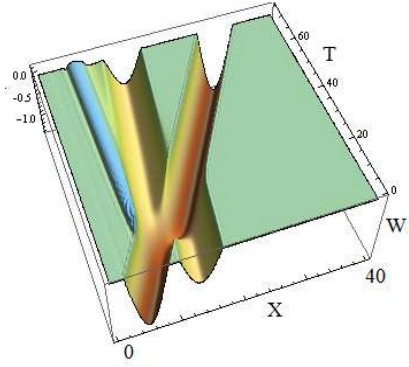}
\end{center}
\caption{Numerical evolution of a pair of  dark compactons   characterized by the velocities $c=1$ and $c=1/4$, respectively.}\label{fig:2}
\end{figure}

To derive a finite-difference scheme, say, for the model equation (\ref{PDE3}), we modify the scheme presented in \cite{frutos}.
In agreement with the methodology proposed in that paper we introduce the artificial viscosity by adding the term $\varepsilon
W_{4x}$, where $\varepsilon$ is a small parameter. Thus, instead of
 (\ref{PDE3}) we have for the case of $n=3$ the following equation:
%  Let $k=1$ i.e.
%\begin{equation}\label{model}
%W_t+\left\{W^3+W\left[W^2\right]_{xx}\right\}_x=0.
%\end{equation}
%
\begin{equation}\label{NestNeq3}
W_t+\left\{W^3\right\}_x+\left\{W\left[W^2\right]_{xx}\right\}_x+\varepsilon
W_{4x}=0.
\end{equation}
Let us approximate the spatial derivatives as follows:
\begin{equation}\label{descrete}
\begin{split}
\frac{1}{120}(\dot W_{j-2}+26\dot W_{j-1}+66\dot W_j+26\dot
W_{j+1}+\dot W_{j+2})+
\\+\frac{1}{24h}(-W_{j-2}^3-10 W_{j-1}^3+10 W_{j+1}^3+W_{j+2}^3)+\\
+\frac{1}{24h}(-L_{j-2}-10 L_{j-1}+10 L_{j+1}+L_{j+2})+ \\
+\varepsilon \frac{1}{h^4}(W_{j-2}-4 W_{j-1}+6W_j
-4W_{j+1}+W_{j+2})=0,
\end{split}
\end{equation}
where $L_j=W_j\frac{W_{j-2}^2-2W_j^2+W_{j+2}^2}{h^2}$ .

To integrate the system (\ref{descrete}) in time, we use the midpoint
method.
%According to this method,
Then the quantities $W_j$ and $\dot
W_j$ are represented in the form
$$ W_j\rightarrow \frac{W_j^{n+1}+W_j^{n}}{2}, \dot W_j\rightarrow
\frac{W_j^{n+1}-W_j^{n}}{\tau}.$$ The resulting nonlinear
algebraic system with respect to  $W_j^{n+1}$ can be solved by
iterative methods.

%We tested the scheme by comparing the solution of the initial value problem with the single comacton %solution taken as a Cauchy data with that obtained by the application of the Galerkin numerical scheme %taken from \cite{Dubin}, Ch. and adapted for equation (\ref{NestNeq3}). The result of the %comparison are shown in figure \ref{fig:GalFrutos}. Thus, we see that the results of the numerical 5experiments are almost identical

We test the scheme (\ref{descrete}) by considering the movement of a
single compacton.  Assume that  the model parameters $c=1$
 and the scheme parameters $N=600$, $h=30/N$, $\tau=0.01$,
$\varepsilon=10^{-3}$ are fixed. The application of the scheme
(\ref{descrete}) gives us fig.\ \ref{fig:1}a.

The starting profile providing the initial condition for the numerical
scheme is chosen according to (\ref{comp2a}) where $n=3$, $c_1=1$ and $c_2=1/4$, namely,
\begin{equation*}
W_{1,2}=\Bigl\{
\begin{array}{l}
\epsilon\sqrt{2c_{1,2}}\cos\left((z-z_{1,2})/2\right) \mbox{if} \left|(z-z_{1,2})/2\right|<\pi/2,\\
0\ \mbox{otherwise}
\end{array}
\end{equation*}
where $z_1=5$, $z_2=13$, and $\epsilon=+1$ for fig.~\ref{fig:1} while $\epsilon=-1$ for fig.~\ref{fig:2}
(note that $W_i$ corresponds to $i$-th figure).

To study the interaction of two bright compactons,
we combine the compacton having the velocity $c=1$ with the slow one characterized by the velocity $c=1/4$ and being shifted to the right at the initial moment of time.
 The result of modelling
is presented at fig.~\ref{fig:1}b. The interaction of two dark compactons has similar properties and is depicted at fig.~\ref{fig:2}.

% CHANG WITH TGE INTERACTION OF DARK-DARK PAIR!!!!!!!!!!!!!!!!!!!!!!!!!!
%Most intriguing case is the interaction of bright and dark
%compactons. According to the papers [], studies of these phenomena
%encounter significant difficulties, mainly, due to the appearance
%of gapes in the zones of  interaction of waves.

\begin{figure}
\begin{center}
\includegraphics[totalheight=1.4 in]{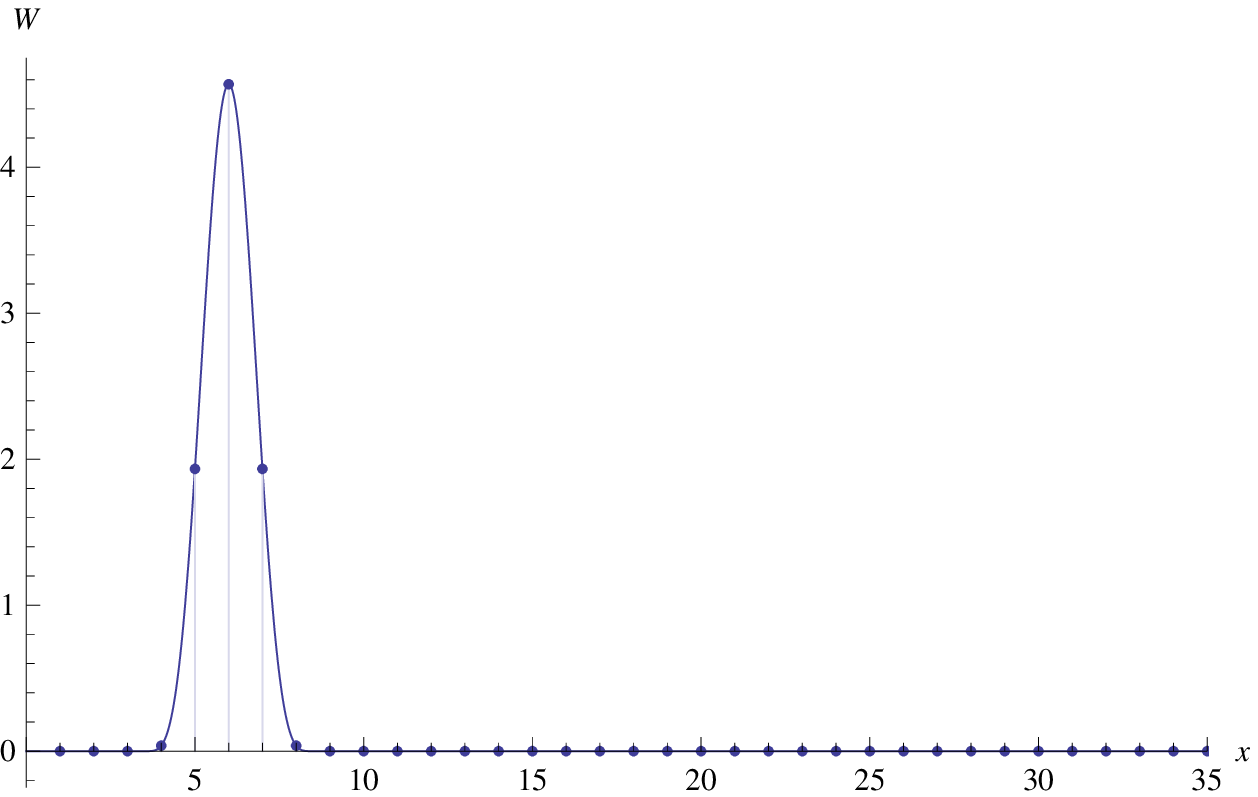}
\includegraphics[totalheight=1.4 in]{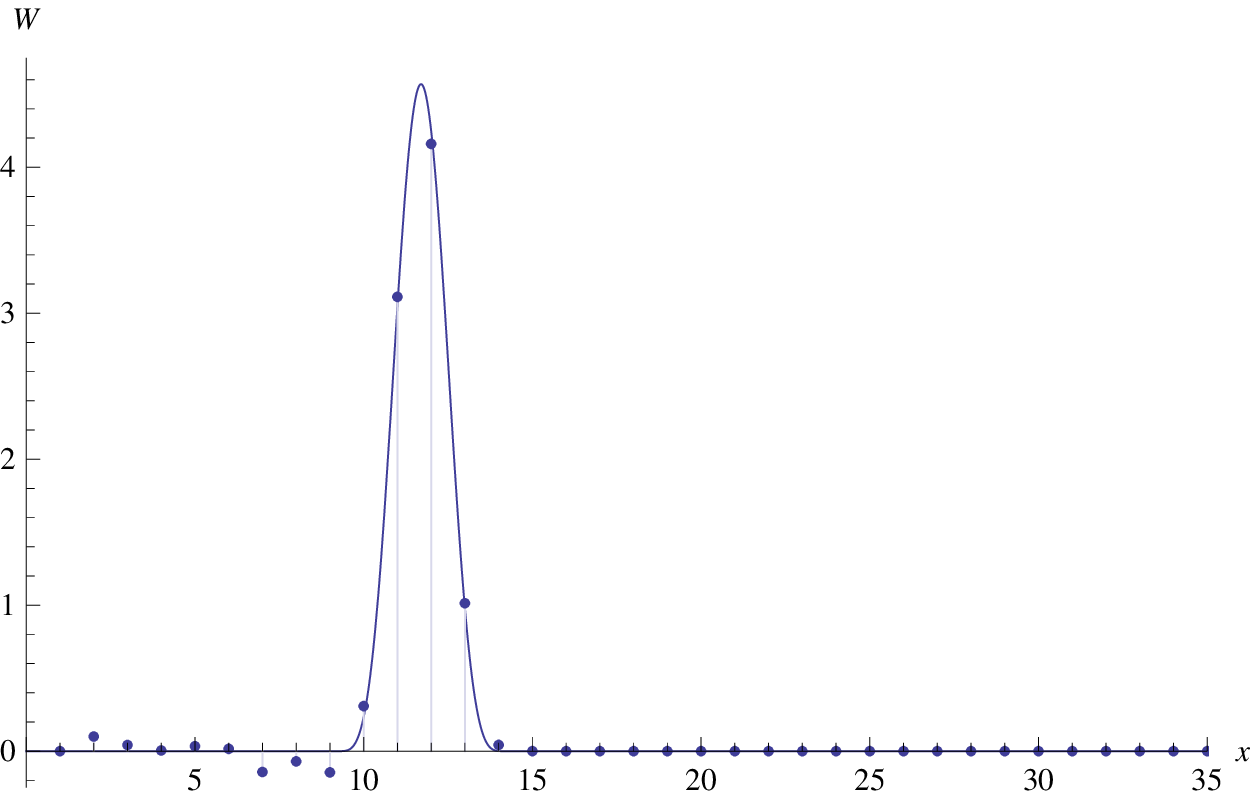}
\\
\includegraphics[totalheight=1.4 in]{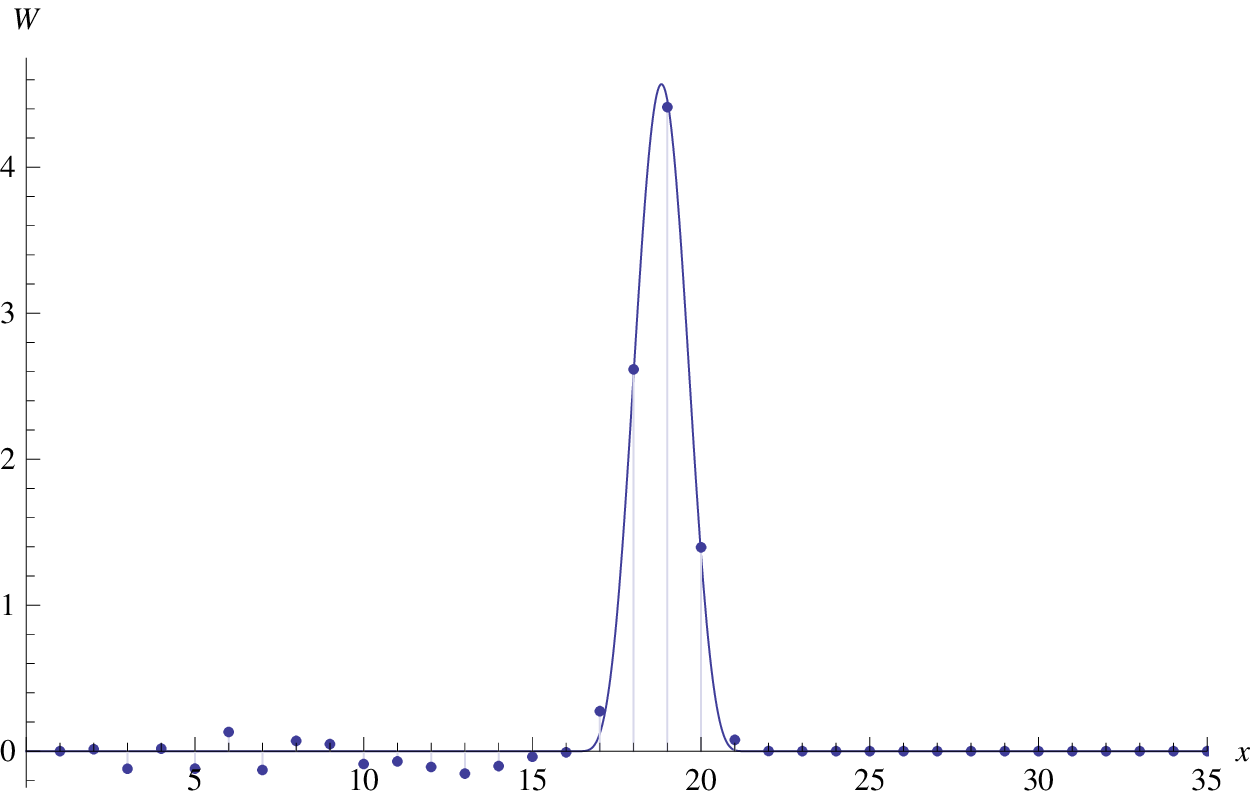}
\includegraphics[totalheight=1.4 in]{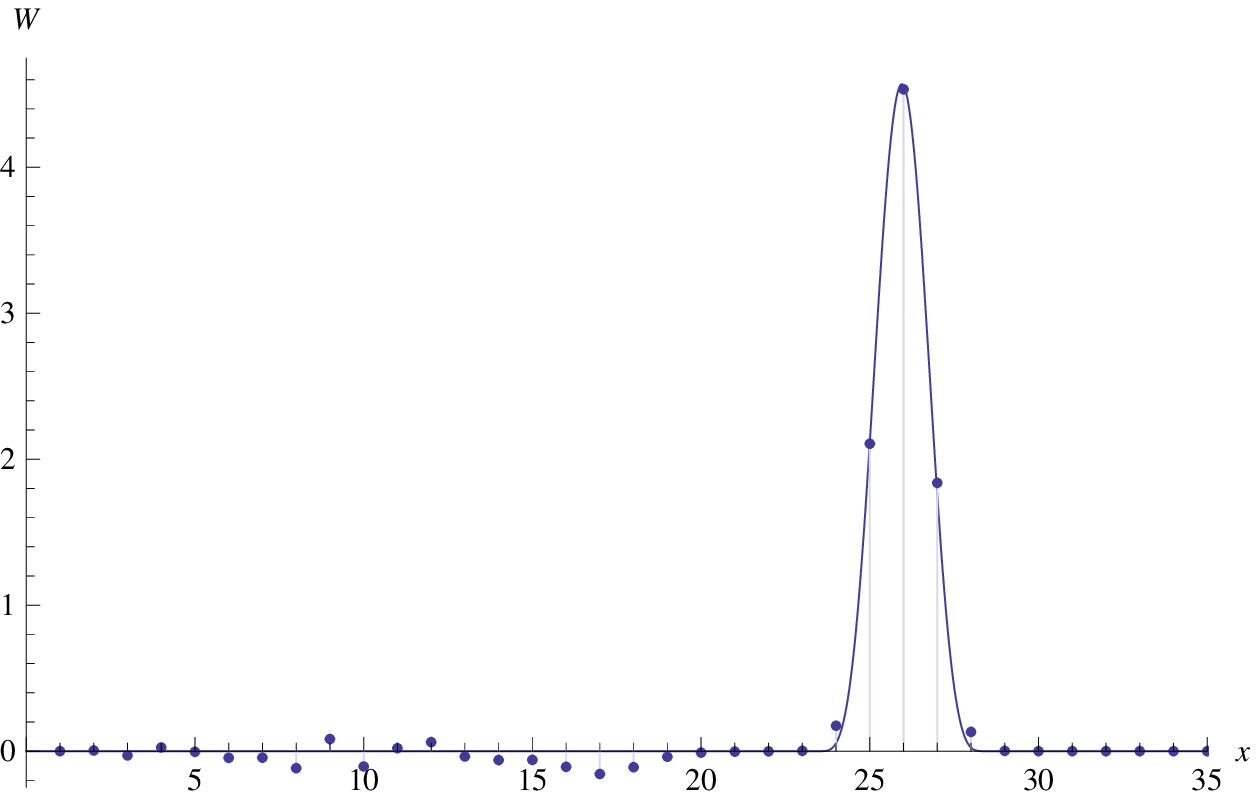}
\caption{Evolution of the initial perturbation in the granular media (marked with dots) on the
 background of the corresponding evolution of the  compacton (marked with solid lines) obtained at the following values of the parameters: $n=3/2$, $c=1.425$, $A=0.25,$ $B=0.3.$  Upper row: left: $t=0;$ right: $t=4$; lower row: left: $t=9$; right: $t=14$
}\label{Fig:singcomp} \end{center}
\end{figure}

%\section{Comparison of numerical evolution of compactons with the numerical solution of the granular %media, subjected to similar initial conditions}

%\noindent
As we have already mentioned at the end of Section \ref{vs:sec2}, there is no way of selecting the scales in the model equations (\ref{PDE3}),   (\ref{PDErar}) and (\ref{PDEComprar}), so the  scaling decomposition employed there is rather formal. Nevertheless, it leads to  interesting equations possessing localized solutions with solitonic features.

Now we are going to compare the evolution of the compacton solutions with {corresponding} solutions of the finite (but long enough) discrete system.
Since the average distance $a$ between adjacent blocks does not play the role of a small parameter anymore, we assume from now on that it is equal to one.
%Upon having made such an assumption,
With this assumption in mind, we can write equation (\ref{PDEComprar}) in the initial variables $t, x$ as follows:
\begin{equation}\label{PDEComprar_tx}
W_{ t}+Q\left[\mathrm{sgn}(W)\right]^{n+1} \left\{W^n+ \hat\beta W^\frac{n-1}{2} \left[W^\frac{n+1}{2}  \right]_{xx}  \right\}_{x}=0,
\end{equation}
where
\[
Q=\frac{A}{\gamma}, \qquad \hat\beta=\frac{n}{6 (n+1)}.
\]
It is easy to verify that equation (\ref{PDEComprar_tx}) possesses the following compacton solutions:
\begin{equation}\label{comp2atx}
W_c^\epsilon(z)=\epsilon W_c(z)=\begin{cases}\epsilon  \tilde M \cos^{\gamma}{\left(\tilde B z\right)}, & \mbox{   if   }  |\tilde B z|<\frac{\pi}{2}, \\
0   & \mbox{  otherwise},\\
\end{cases}
\end{equation}
where $\epsilon = \pm 1,$  $z=x-c t,$
\[
 \tilde M=\left[\frac{c (n+1)}{2 Q}  \right]^{\frac{1}{n-1}}, \qquad \tilde B=\frac{n-1}{(n+1) \sqrt{\beta}},  \qquad \gamma=\frac{2}{n-1}.
\]

\noindent
We introduce the functions $R_k=Q_{k-1}-Q_k$ being the discrete analogs to the strain field $W(t, x)$. These functions are assumed to satisfy the system
\begin{equation}\label{DiscrSys}
\begin{array}{rcl}
\ddot R_1(t)&=&0, \\
\ddot R_k(t)&=&A\left[ R_{k-1} |R_{k-1}|^{n-1}-2 R_{k} |R_{k}|^{n-1}+R_{k+1} |R_{k+1}|^{n-1} \right]\\[2mm]
&&+\gamma \left[ R_{k-1} |R_{k-1}|^{n-1}-2 R_{k} |R_{k}|^{n-1}+R_{k+1} |R_{k+1}|^{n-1} \right],\\[2mm]
k&=&2,\dots,m-1, \\
\ddot R_m(t)&=&0
\end{array}
\end{equation}
We solve this system with the following initial conditions induced by the compacton solution (\ref{comp2atx}) in the respective nodes:
\begin{equation}\label{discrinit}
R_k(0) = \begin{cases}   \epsilon \tilde M \cos^\gamma [ \tilde B   k-I]  & \mbox{if}\ |\tilde B   k-I|<\pi/2  \\
0 & \mbox{otherwise},  \end{cases}
\end{equation}
\begin{equation}\label{primdiscrinit}
\dot R_k(0) = \begin{cases}    \epsilon \tilde M c \gamma \tilde B \cos^{\gamma-1} [ \tilde B   k-I]\sin[ \tilde B    k- I]  &\mbox{if } |\tilde B k-I|<\pi/2  \\
0 & \mbox{  otherwise},  \end{cases}
\end{equation}
\begin{equation}
R_1(0)=\dot R_1(0)=R_m(0)=\dot R_m(0)=0,
\end{equation}
where $I$ is a constant phase, $k=2,3,\dots,m-1$.
Note that  $A$ and $\gamma$  appear in equation (\ref{PDEComprar_tx}) in the form of the  ratio $Q=A/\gamma$, whereas in the system (\ref{DiscrSys})
they appear as independent parameters. Therefore, one
should not expect a one-to-one correspondence between the solutions of the discrete and continuous problems for arbitrary values of the
parameters. The numerical experiments confirm this hypothesis by showing that synchronous evolution of the
same compacton perturbation within two models can be observed for a unique value of the velocity
$c=c_{0}$. This value depends strongly on the parameter $\gamma$ and depends on the parameter $A$ in a much weaker fashion.  It has been observed that at $c<c_0$ the discrete compacton moves
quicker than its continuous analogue while at $c>c_0$  the opposite effect occurs.
The result of comparison for a single compacton is shown at fig.~\ref{Fig:singcomp}. One  can see
that at the chosen values of the parameters the main perturbations  move synchronously and do not
change their form. However, in the tail part of the  discrete analogue small nonvanishing oscillations appear after a while.

\begin{figure}
\begin{center}
\includegraphics[totalheight=1.4 in]{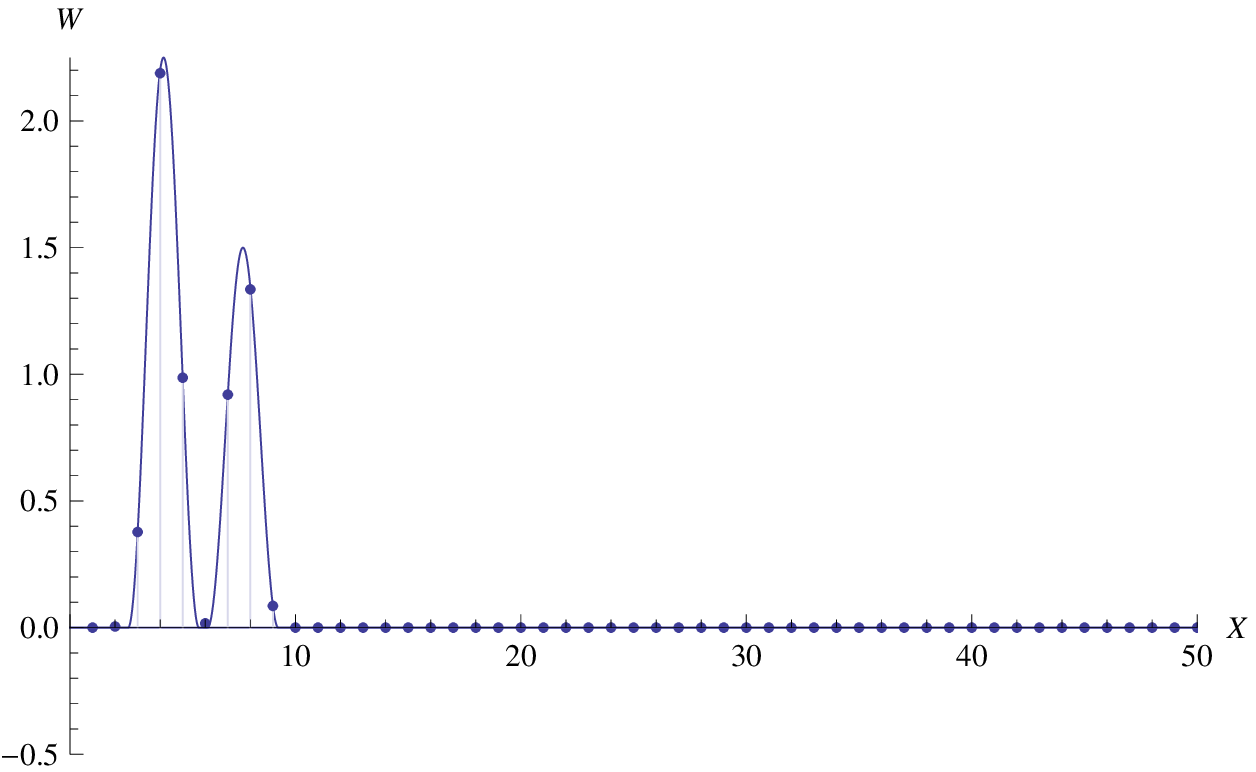}
\includegraphics[totalheight=1.4 in]{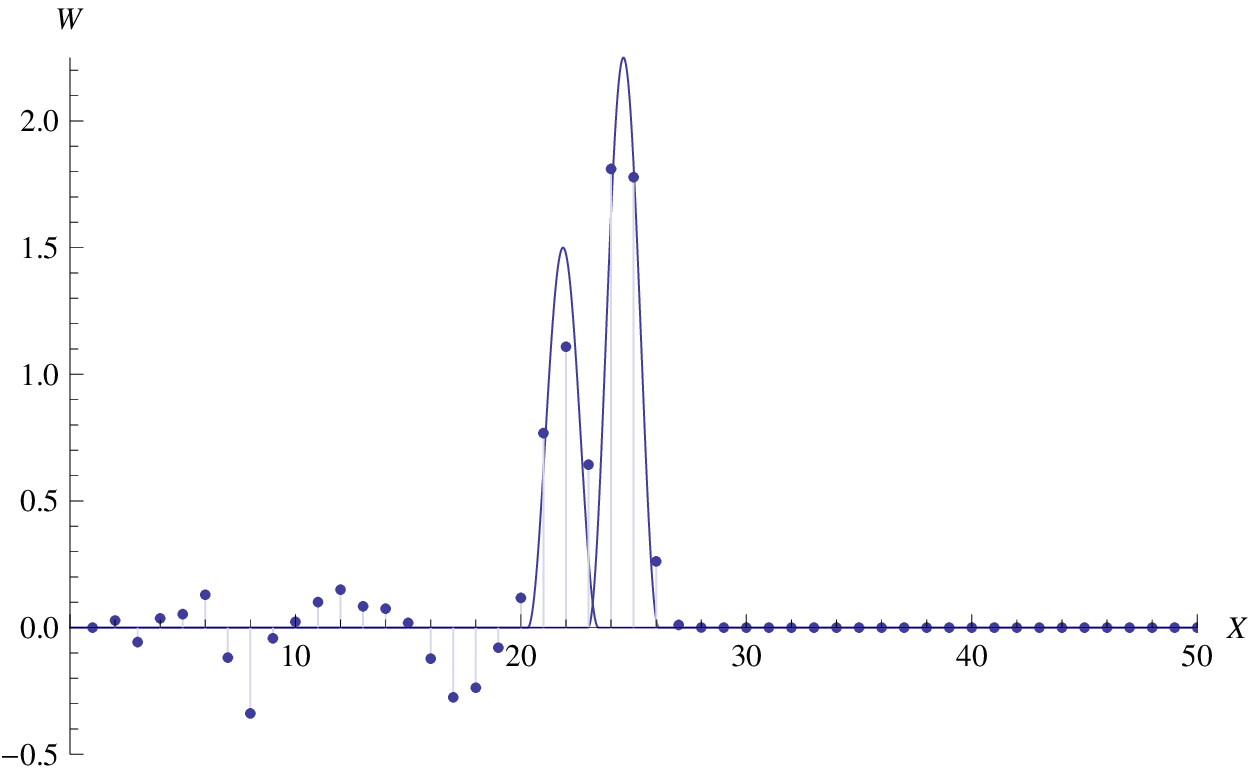}
\\
\includegraphics[totalheight=1.4 in]{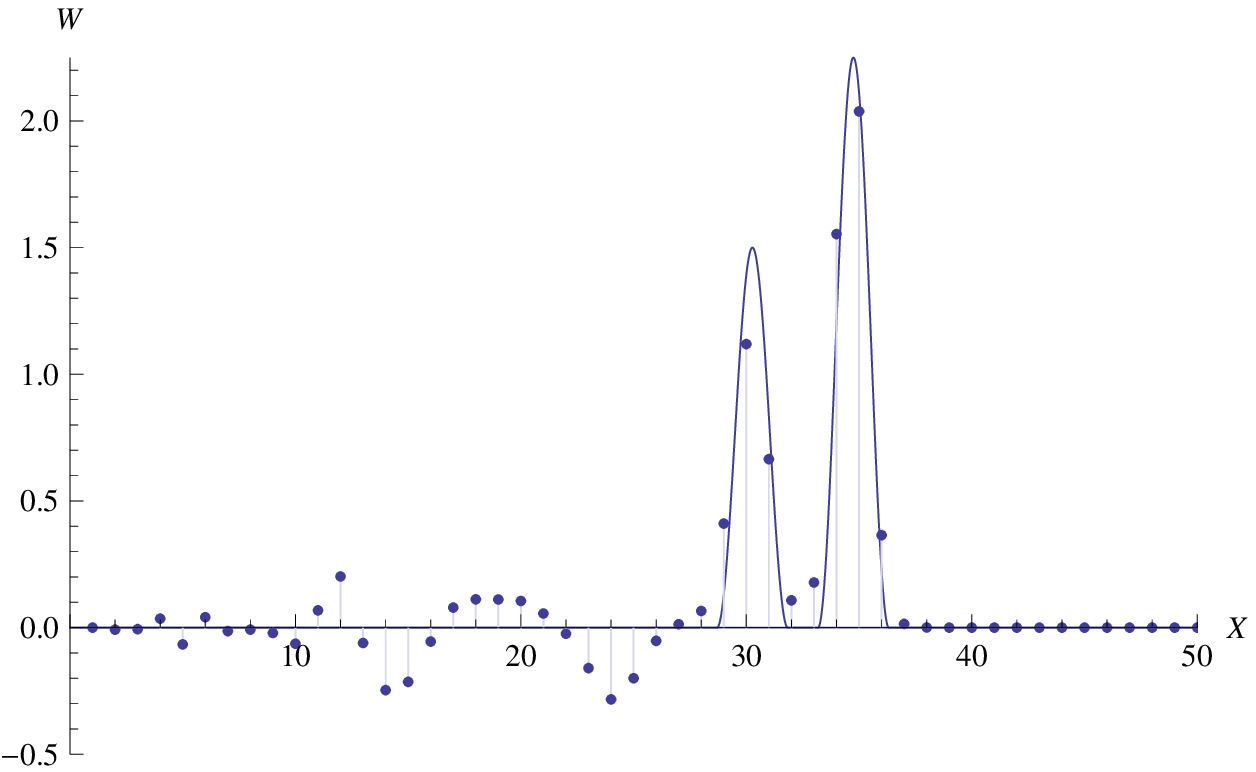}
\includegraphics[totalheight=1.4 in]{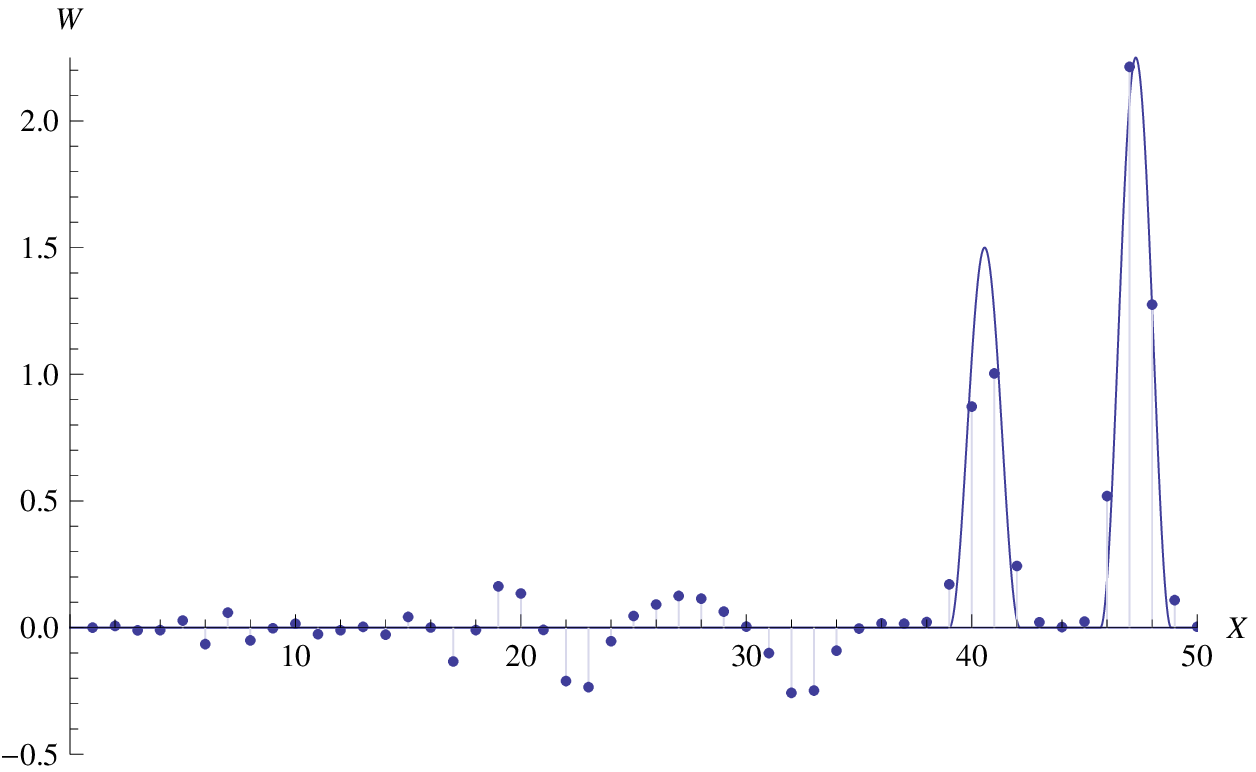}
\caption{Evolution of two initially separated compacton perturbation in the granular media (marked with dots) on the background of the corresponding compacton solutions of the continual model   (marked with solid lines), obtained at the following values of the parameters: $n=2$, $c_1=1.5$, $c_2=1.0$, $A=B=1.$ Upper row: left: $t=0$; right: $t=12$; lower row: left: $t=18.25$; right: $t=26$
}\label{Fig:collisions} \end{center}
\end{figure}

Since for every value of the parameter $\gamma$ there is a unique value of the wave pack velocity for
which the discrete and continuous compacton perturbations move synchronously, one should not expect
that the collision of compactons within these two models will proceed in the same way for any
set of values of parameters. However, collision processes display not much of qualitative differences for the  discrete
pulses which interact elastically like their continuous analogues. This is illustrated on
fig.~\ref{Fig:collisions} showing the evolution of two initially separated discrete compactons. For
convenience,  the continuous compactons which coincide with the right-hand side of the initial data (\ref{discrinit}) at $t=0$ (the leftmost graph in the first row) are also shown in this figure. Continuous curves shown on the following graphs are obtained by appropriate translations.  They are presented in order to emphasize the quasi-elastic nature of interaction of the discrete pulses.

\section{Conclusions and discussion}  \label{vs:sec5}

%\noindent

In the present paper we have studied compacton solutions supported by the nonlinear evolutionary PDEs. The equations we considered, (\ref{PDE3}), (\ref{PDErar}), and (\ref{PDEComprar}), are obtained from the  dynamical system (\ref{DS_Nest}) describing one-dimensional chain of prestressed elastic bodies. Equation (\ref{EqNest2}) obtained in \cite{Nester_02} from this model without resorting to the method of multi-scaled decomposition possesses the compacton solutions which fail the stability test. Numerical simulations show that the compacton solutions supported by equation (\ref{EqNest2}) are destroyed in a very short time.

In contrast with the above, equations  (\ref{PDE3}) (resp.\ (\ref{PDErar})), which are obtained using formal multiscale decomposition,
possess families  of bright (resp.\ dark) compacton solutions which appear to be stable. This is backed both by the stability test and the results of the numerical simulations.

As we have shown in Sections \ref{sec:hs}--\ref{sec:int}, for generic values of the parameter $n$ equation (\ref{PDE3}) does not possess an infinite set of higher symmetries or other signs of complete integrability such as infinite hierarchies of conservation laws.  Nevertheless the compacton solutions to this equation possess some features which  are characteristic for ``genuine"  soliton solutions. In this connection it would be interesting to compare the traveling wave solutions for the distinguished case $n=-2$ with any other  equation of the family (\ref{PDE3}) with negative $n$.
Qualitative analysis of the factorized equations describing the TW solutions shows that there are no
compacton solutions for the models with the negative $n$, but nevertheless all of them seem to possess periodic solutions resembling peakons. It would be interesting to find out whether there is any difference in the qualitative behavior of periodic solutions of the only integrable case ($n=-2$) in comparison with the periodic TW solutions supported by the model characterized by other values $n<0$. Perhaps the differences will be manifested in the stability  properties as this is the case with the soliton solutions supported by the family of the KdV-type equations.\looseness=-1

A characteristic feature of equations (\ref{PDE3}), (\ref{PDErar}) related to the decomposition we used is that they describe processes
with ``long" temporal and  ``short" spatial scales. Hence it is rather questionable whether these equations can adequately describe a localized pulse propagation in discrete media in the situation when the distance between the adjacent  particles is comparable to the compacton width $\Delta x$. In fact, making the ``reverse'' transformations $X\rightarrow\xi\rightarrow x$ we get the following formula for the width of the compacton solution (\ref{comp2a}) in the initial coordinate system:
\[
\Delta x= \pi a \sqrt{\frac{n (n+1)}{6 (n-1)^2}};
\]
this is nothing but equation (1.130) from \cite{Nester_02}.
For $n=3/2$, corresponding to the Hertzian force between spherical particles, we get $\Delta x\approx 4.96 a$. It is then interesting to notice that the same results for the particles with the spherical geometry were obtained in the course of numerical simulations,  and experimental studies \cite{Nester_83, NestLaz_85, Nester_94,Nester_95,Vengr}. We wish to stress that  results of our analysis as well as the main conclusions are in agreement with the earlier publications by other authors. In particular,  P. Rosenau notes, when  considering the general models of dense chains \cite{Rosenau_06}, that the natural separation of scales leading to an unidirectional PDE of first order in time does not exist.

%In closing note that some of the present results, in particular those concerning the stability study, are of somewhat preliminary nature.
%The full investigation of stability of compacton solutions supported by equations (\ref{PDE3}), (\ref{PDErar}), and more general equation %(\ref{PDEComprar}), will be published elsewhere.

\section*{Acknowledgments}
VV gratefully acknowledges support from the Polish Ministry of Science
and Higher Education. The research of AS was supported in part by the RVO funding for
I\v{C}47813059, and by the Grant Agency of the Czech Republic (GA \v{C}R)
under grant P201/12/G028. AS gratefully acknowledges warm hospitality extended to him in the course of his visits to AGH in Krak\'ow.\looseness=-1

We are pleased to thank the anonymous referee and M.V. Pavlov for useful suggestions.
%\newpage
%\bibliographystyle{siamplain}
%\bibliography{references}

\end{document}